\documentclass[12pt,epsf,fleqn]{article}
\usepackage{epsfig}
\usepackage[dvips]{color}
\begin{document}
\title{Neutral scalar Higgs bosons in the USSM at the LHC}
\author{S. W. Ham$^{(1)}$, Taeil Hur$^{(2)}$, P. Ko$^{(3)}$, and S. K. Oh$^{(4)}$
\\
\\
{\it $^{(1)}$ Center for High Energy Physics, Kyungpook National University}
\\
{\it Daegu 702-701, Korea}
\\
{\it $^{(2)}$ Department of Physics, KAIST, Daejon 305-701, Korea}
\\
{\it $^{(3)}$ School of Physics, KIAS, Seoul 130-722, Korea}
\\
{\it $^{(4)}$ Department of Physics, Konkuk University, Seoul 143-701, Korea}
\\
\\
}
\date{}
\maketitle
\begin{abstract}
We study the possibility of discovering neutral scalar Higgs bosons
in the $U(1)'$-extended supersymmetric standard model (USSM) at the CERN Large Hadron Collider (LHC),
by examining their productions via the exotic quark loop in the gluon fusion process at leading order.
It is possible in some parameter region that the neutral scalar Higgs bosons may have stronger couplings
with the exotic quarks than with top quark.
In this case, the exotic quarks may contribute more actively than top quark
in productions of the neutral scalar Higgs bosons in the gluon fusion process.
We find that there is indeed some parameter region in the USSM that support our speculations.
\end{abstract}
\vfil\eject

%***********************************************************************
\section{INTRODUCTION}
%***********************************************************************

Enlarging the Higgs sector of the Standard Model (SM) and studying its phenomenology have been one
of the hot subjects for both theoretical and experimental physicists.
The primary motivation to enlarge the SM Higgs sector is to solve the naturalness problem,
which is essentially how to avoid the quadratic divergence in radiative corrections
to the Higgs boson mass arising from the SM particle loops.
One of the most attractive solutions to the naturalness problem is provided by the supersymmetry (SUSY),
in which the loop corrections of the SM particles are cancelled
by the loop corrections of their superpartners [1-3].
In this context, the SUSY is regarded as a gateway to new physics at TeV scale.

Among a number of supersymmetric extensions of the SM,
the simplest one is the minimal supersymmetric standard model (MSSM),
which possesses only two Higgs doublets.
The superpotential of the MSSM should contain a mixing term between the two Higgs doublets,
proportional to $\mu$ of TeV or lower scale,
in order to generate the right size of the electroweak symmetry breaking.
The parameter $\mu$ has mass dimension, sometimes regarded as a drawback of the MSSM [4].
This is the so-called $\mu$ problem of the MSSM, which may be solved by introducing an extra Higgs singlet
in addition to the two Higgs doublets of the MSSM [5,6].
A number of nonminimal supersymmetric standard models have at least one Higgs singlet in their Higgs sectors,
where the parameter $\mu$ of the MSSM is replaced in terms of the vacuum expectation value (VEV)
of the Higgs singlet field.

These nonminimal supersymmetric standard models may be divided into two classes according to
whether their gauge groups contain an additional $U(1)'$ or not.
A representative example of those models that have no additional $U(1)'$ may be
the next-to-minimal supersymmetric standard model (NMSSM) [7-15].
The minimal non-minimal supersymmetric standard model (MNMSSM) also has no additional $U(1)'$,
where the global $U(1)$ Peccei-Quinn symmetry is explicitly broken by means of the tadpole term of the Higgs singlet [16-20].
On the other hand, there is the $U(1)'$-extended supersymmetric standard mode (USSM)
as a typical example of those models that have additional $U(1)'$ [21-38].

The USSM may emerge from the superstring-inspired $E_6$ model,
which has two additional $U(1)$ symmetries besides the SM gauge group.
These two $U(1)$ symmetries are linearly mixed to yield the desired $U(1)'$,
where the mixing angle is denoted as $\theta_E$.
For $\theta_E=0$, the USSM is called as the $\chi$-model.
Similarly, it is called the $\psi$-model for $\theta_E=\pi/2$,
the $\eta$-model for $\tan^{-1} (-\sqrt{5/3})$, and the $\nu$-model for $\tan^{-1} (\sqrt{15})$.
A characteristic of the superstring-inspired $E_6$ model is that
it possesses naturally an extra pair of $SU(2)$ singlet quarks, $D_k$ and ${\bar k}_R$,
with electric charge $-1/3$ and $+1/3$, respectively.
These exotic quarks are introduced in order to fill in the fundamental 27 representation of $E_6$.
They may participate in the USSM phenomenology, if they are sufficiently light.

At the Large Hadron Collider (LHC), the dominant production mechanism
for the SM Higgs boson with a mass below 1 TeV is the gluon fusion process
which is mediated through a top quark loop [39-42].
This production depends directly on the coupling of the Higgs boson to a top quark pair.
In the same manner, the light exotic quarks of the USSM may take part in the Higgs productions at the LHC.

We study in this article the Higgs productions of the USSM at the LHC
via gluon fusion process through the exotic quark loop at the leading order.
The masses and the mixing matrix of the neutral scalar Higgs bosons are calculated
using the effective potential approximation at the one-loop level,
by considering radiative corrections due to the top quarks, bottom quarks,
exotic quarks and their superparticles.
For the gluon fusion process, both the top quark loop and the exotic quark loop are studied
for the productions of the neutral scalar Higgs bosons in the USSM.
It is found that the lightest neutral scalar Higgs boson in the USSM may couple dominantly
to a top quark pair when the relevant parameters have certain values.
In this case, the lightest scalar Higgs boson would produce dominantly
through the gluon fusion process via a top-quark loop at the LHC.
At the same time, one of two heavier scalar Higgs bosons couples strongly
to an exotic-quark pair because of an orthogonality of the transformation matrix.
In this scenario, the gluon-fusion process via an exotic-quark loop
on the production of the heavier scalar Higgs boson is more dominant mechanism
than that via a top-quark loop.

%***********************************************************************
\section{THE HIGGS SECTOR}
%***********************************************************************

The Higgs sector of the USSM consist of two Higgs doublets, $H_1 = (H_1^0, H_1^-)$
and $H_2 = (H_2^+, H_2^0)$, and a neutral Higgs singlet, $S$.
There are therefore ten real degrees of freedom.
For simplicity, we take only the third generation of quarks into account.
Then, the superpotential for the Yukawa interactions in the USSM for quarks
and the exotic quarks may be expressed as
\begin{equation}
W \approx h_t Q^T \epsilon H_2 t_R^c + h_b Q^T \epsilon H_1 b_R^c + h_k S k_L {\bar k}_R
+ \lambda H_1^T \epsilon H_2 S  \ ,
\end{equation}
where $\epsilon$ is an antisymmetric $2 \times 2$  matrix with $\epsilon_{12} = 1$,
and $h_t$, $h_b$ and $h_k$ are respectively the dimensionless Yukawa coupling coefficients of
top, bottom, and exotic quarks,
$t_R^c$ and $b_R^c$ are the right-handed top and bottom quark superfields, respectively,
$Q$ is the left-handed $SU(2)$ doublet quark superfield of the third generation,
and the right and left handed singlet exotic quark superfields are denoted respectively as $k_R$ and $k_L$.

The Higgs potential at the tree level may be collected from the on-shell Lagrangian as
\begin{equation}
    V_0 = V_F + V_D + V_{\rm S} \ ,
\end{equation}
where
\begin{eqnarray}
V_F & = & |\lambda|^2 [(|H_1|^2 + |H_2|^2) |S|^2 + |H_1^T \epsilon H_2|^2]  \ , \cr
V_D & = & {g_2^2 \over 8} (H_1^{\dagger} \vec\sigma H_1 + H_2^{\dagger} \vec\sigma H_2)^2
+ {g_1^2 \over 8} (|H_1|^2 - |H_2|^2)^2 \cr
& &\mbox{}+ {g_1'^2 \over 2} ( {\tilde Q}_1 |H_1|^2
+ {\tilde Q}_2 |H_2|^2 + {\tilde Q}_3 |S|^2)^2 \ , \cr
V_{\rm S} & = & m_1^2 |H_1|^2 + m_2^2 |H_2|^2 + m_3^2 |S|^2
- [\lambda A_{\lambda} (H_1^T \epsilon H_2) S + {\rm H.c.}] \ ,
\end{eqnarray}
where $\vec \sigma$ denotes the three Pauli matrices,
$g_2$, $g_1$, and $g_1'$ are the $SU(2)$, $U(1)_Y$, and $U(1)'$ gauge-coupling constants, respectively,
${\tilde Q}_1$, ${\tilde Q}_2$, and ${\tilde Q}_3$ are the effective $U(1)'$ hypercharges
of  $H_1$, $H_2$, and $S$, respectively.
These three effective $U(1)'$ hypercharges should satisfy $\sum_{i=1}^3 {\tilde Q}_i = 0$,
since the superpotential should be invariant under $U(1)'$.
The Higgs potential also introduces three soft masses, $m_i$ ($i=1,2,3$),
which can eventually be eliminated by requiring the minimum conditions for the VEVs.
The VEVs of the three neutral Higgs fields are denoted as $v_1 = <H_1^0>$, $v_2 = <H_2^0>$, and $s = <S>$.
The parameters of the Higgs potential as well as the VEVs of the three neutral Higgs fields
are assumed to be real, since we do not consider the CP mixing
between the scalar and pseudoscalar Higgs bosons in this article.

At the tree level, the masses of stop, sbottom,
and exotic squarks (${\tilde t}_1, {\tilde t}_2, {\tilde b}_1, {\tilde b}_2, {\tilde k}_1, {\tilde k}_2$)
are respectively given by
\begin{eqnarray}
m_{{\tilde t}_1, \ {\tilde t}_2}^2 & = & m_Q^2 + m_t^2 \mp m_t
|A_t - \lambda s \cot \beta| \ , \cr
m_{{\tilde b}_1, \ {\tilde b}_2}^2 & = & m_Q^2 + m_b^2 \mp m_b
|A_b - \lambda s \tan \beta| \ , \cr
m_{{\tilde k}_1, \ {\tilde k}_2}^2 & = & m_K^2 + m_k^2 \mp m_k
|A_k - \lambda v^2 \sin 2 \beta/(2 s)| \ ,
\end{eqnarray}
where $\tan\beta = v_2/v_1$ and $v^2 = v^2_1 + v^2_2$.
Also, the masses of top, bottom, and the exotic quarks ($t,b,k$) at the tree level are
respectively given by $m_t = h_t v_2$, $m_b = h_b v_1$, and $m_k = h_k s$.
Note that $g_1'$ is generally the order of $g_1$, and the $V_D$ does not contribute to the squark masses.
We assume that the soft SUSY-breaking masses for the left and right squarks are same.

The Higgs potential of the USSM at the one-loop level may be obtained
by including the radiative corrections due to quarks and squarks of the third generation,
and the exotic quarks and squarks.
In general, the radiative corrections due to the exotic quarks and squarks can
be significant since the Yukawa couplings of the exotic quarks to the Higgs bosons can be
large at the electroweak scale.
The one-loop effective potential is given as [43]
\begin{equation}
V_1  = \sum_{l} {n_l {\cal M}_l^4 \over 64 \pi^2}
\left [
\log {{\cal M}_l^2 \over \Lambda^2} - {3 \over 2}
\right ]  \ ,
\end{equation}
where $\Lambda$ is the renormalization scale in the modified minimal subtraction scheme,
and the subscript $l$ stands for $t$, $b$, $k$ as well as  ${\tilde t}_1$, ${\tilde t}_2$,
${\tilde b}_1$, ${\tilde b}_2$, ${\tilde k}_1$, and ${\tilde k}_2$.
The degrees of freedom for these quarks and squarks are $n_t = n_b = n_k = -12$
and $n_{{\tilde t}_i} = n_{{\tilde b}_i} = n_{{\tilde k}_i} = 6$ ($i=1,2$).
Therefore, the full Higgs potential at the one-loop level is given as $V = V_0 + V_1$.

%***********************************************************************
\subsection{HIGGS MASS}
%***********************************************************************

The ten real degrees of freedom of the Higgs sector of the USSM is linearly combined
to yield two neutral Goldstone bosons, a pair of charged Goldstone bosons,
four neutral Higgs bosons and a pair of charged Higgs bosons.
After the electroweak symmetry breaking, the four Goldstone bosons, neutral and charged,
would eventually be absorbed into the longitudinal component of $Z$, $Z'$, and $W$ gauge bosons,
where $Z'$ is the additional neutral gauge boson of the USSM.
Note that we have definite CP parities for the neutral Higgs bosons.
Thus, as physical Higgs bosons, the USSM has one neutral pseudoscalar Higgs boson,
three neutral scalar Higgs bosons, and a pair of the charged Higgs bosons.

At the one-loop level, the squared mass of the pseudoscalar Higgs boson is obtained as
\begin{eqnarray}
m_A^2 & = & {2 \lambda v \over \sin 2 \alpha} \left [A_{\lambda}
- {3 m_t^2 A_t \over 16 \pi^2 v^2 \sin^2 \beta} f (m_{{\tilde t}_1}^2,  \ m_{{\tilde t}_2}^2)
\right. \cr
& &\mbox{} \left.
- {3 m_b^2 A_b \over 16 \pi^2 v^2 \cos^2 \beta} f (m_{{\tilde b}_1}^2,  \ m_{{\tilde b}_2}^2)
- {3 m_k^2 A_k \over 16 \pi^2 s^2} f (m_{{\tilde k}_1}^2,  \ m_{{\tilde k}_2}^2) \right ] \ ,
\end{eqnarray}
where the dimensionless function $f(m_x^2, \ m_y^2)$, arising from radiative corrections, is defined as
\[
 f(m_x^2, \ m_y^2) = {1 \over (m_y^2 - m_x^2)} \left[  m_x^2 \log {m_x^2 \over \Lambda^2} - m_y^2
\log {m_y^2 \over \Lambda^2} \right] + 1 \ ,
\]
and the mixing angle $\alpha$, standing for the splitting
between an extra $U(1)'$ symmetry breaking scale and the electroweak scale, is given as
\[
    \tan \alpha = {v \over 2 s} \sin 2 \beta \ .
\]

The mass matrix $M$ for the three neutral scalar Higgs bosons at the one-loop level is expressed
by a symmetric real $3\times 3$ matrix as
\begin{equation}
M = \left ( \begin{array}{ccc}
    M_{11} & M_{12} & M_{13}   \cr
    M_{12} & M_{22} & M_{23}   \cr
    M_{13} & M_{23} & M_{33}
        \end{array}
    \right ) \ ,
\end{equation}
where its elements may conveniently be decomposed as
\[
    M_{ij} = M_{ij}^0 + M_{ij}^t + M_{ij}^b + M_{ij}^k
\]
where $M_{ij}^0$ is the matrix elements of the mass matrix at the tree level,
obtained from $V^0$, and $M_{ij}^t$, $M_{ij}^b$, and $M_{ij}^k$ are
respectively the one-loop contributions from the top quark sector, the bottom quark sector,
and the exotic quark sector.
They are obtained from $V^1$.

Explicitly, the elements of the mass matrix for the three neutral scalar Higgs bosons
at the tree level are obtained as
\begin{eqnarray}
M_{11}^0 & = & m_Z^2 \cos^2 \beta + 2 g'^2_1 {\tilde Q}_1^2 v^2 \cos^2 \beta
+ m_A^2 \sin^2 \beta \cos^2 \alpha  \ ,  \cr
M_{22}^0 & = & m_Z^2 \sin^2 \beta + 2 g'^2_1 {\tilde Q}_2^2 v^2 \sin^2 \beta
+ m_A^2 \cos^2 \beta \cos^2 \alpha \ ,  \cr
M_{33}^0 & = & 2 g'^2_1 {\tilde Q}_3^2 s^2 + m_A^2 \sin^2 \alpha \ , \cr
M_{12}^0 & = & g'^2_1 {\tilde Q}_1 {\tilde Q}_2 v^2 \sin 2 \beta + (\lambda^2 v^2 - m_Z^2/2) \sin 2 \beta
- m_A^2 \cos \beta \sin \beta \cos^2 \alpha \ ,  \cr
M_{13}^0 & = & 2 g'^2_1 {\tilde Q}_1 {\tilde Q}_3 v s \cos \beta + 2 \lambda^2 v s \cos \beta
- m_A^2 \sin \beta \cos \alpha \sin \alpha \ , \cr
M_{23}^0 & = & 2 g'^2_1 {\tilde Q}_2 {\tilde Q}_3 v s \sin \beta + 2 \lambda^2 v s \sin \beta
- m_A^2 \cos \beta \cos \alpha \sin \alpha  \ .
\end{eqnarray}

From $V^1$, we obtain the one-loop contributions from the top quark sector, $M^t$, as
%*********************  top and stop ******************************
\begin{eqnarray}
    M_{11}^t & = & {3 m_t^4 \lambda^2 s^2 \Delta_{\tilde t}^2 \over 8 \pi^2  v^2 \sin^2 \beta}
{g(m_{{\tilde t}_1}^2, \ m_{{\tilde t}_2}^2) \over (m_{{\tilde t}_2}^2 - m_{{\tilde t}_1}^2)^2}  \ , \cr
%%%%%%%%%
    M_{22}^t & = & {3 m_t^4 A_t^2 \Delta_{\tilde t}^2 \over 8 \pi^2  v^2 \sin^2 \beta}
{g(m_{{\tilde t}_1}^2, \ m_{{\tilde t}_2}^2) \over (m_{{\tilde t}_2}^2 - m_{{\tilde t}_1}^2)^2}
 + {3 m_t^4 A_t \Delta_{\tilde t} \over 4 \pi^2 v^2 \sin^2 \beta}
{\log (m_{{\tilde t}_2}^2 / m_{{\tilde t}_1}^2) \over (m_{{\tilde t}_2}^2 - m_{\tilde{t}_1}^2)} \cr
& &\mbox{} + {3 m_t^4 \over 8 \pi^2 v^2 \sin^2 \beta}
\log \left ( {m_{\tilde t}^2  m_{{\tilde t}_2}^2 \over m_t^4} \right ) \ , \cr
%%%%%%%%%
    M_{33}^t & = & {3 m_t^4 \lambda^2 \Delta_{\tilde t}^2 \over 8 \pi^2 \tan^2 \beta}
{g(m_{{\tilde t}_1}^2, \ m_{{\tilde t}_2}^2) \over (m_{{\tilde t}_2}^2 - m_{{\tilde t}_1}^2 )^2} \ , \cr
%%%%%%%%%
    M_{12}^t & = &\mbox{} - {3 m_t^4 \lambda A_t s \Delta_{\tilde t}^2
\over 8 \pi^2 v^2 \sin^2 \beta}
{g(m_{{\tilde t}_1}^2, \ m_{{\tilde t}_2}^2) \over (m_{{\tilde t}_2}^2 - m_{{\tilde t}_1}^2)^2}
- {3 m_t^4 \lambda s \Delta_{\tilde t} \over 8 \pi^2 v^2 \sin^2 \beta}
{\log (m_{{\tilde t}_2}^2 / m_{{\tilde t}_1}^2) \over (m_{{\tilde t}_2}^2 - m_{{\tilde t}_1}^2)} \cr
%%%%%%%%%
    M_{13}^t & = & {3 m_t^4 \lambda^2 s \Delta_{\tilde t}^2 \over 8 \pi^2 v \sin \beta \tan \beta}
{g(m_{{\tilde t}_1}^2, \ m_{{\tilde t}_2}^2) \over (m_{{\tilde t}_2}^2 - m_{{\tilde t}_1}^2)^2 }
- {3 m_t^2 \lambda^2 s \cos \beta \over 8 \pi^2 v \sin^2 \beta}
f(m_{{\tilde t}_1}^2, \ m_{{\tilde t}_2}^2) , \cr
%%%%%%%%%
    M_{23}^t & = &\mbox{} - {3 m_t^4 \lambda A_t \Delta_{\tilde t}^2
\over 8 \pi^2 v \sin \beta \tan \beta}
{g(m_{{\tilde t}_1}^2, \ m_{{\tilde t}_2}^2) \over (m_{{\tilde t}_2}^2 - m_{{\tilde t}_1}^2)^2}
- {3 m_t^4 \lambda \cos \beta \Delta_{\tilde t} \over 8 \pi^2 v \sin^2 \beta}
{\log (m_{{\tilde t}_2}^2 / m_{{\tilde t}_1}^2) \over (m_{{\tilde t}_2}^2 - m_{{\tilde t}_1}^2) }  ,
\end{eqnarray}
where
\begin{eqnarray}
 \Delta_{\tilde t} & = & A_t - \lambda s \cot \beta  \  ,
\end{eqnarray}
and the dimensionless function $g$ is defined as
\[
 g(m_x^2,m_y^2) = {m_y^2 + m_x^2 \over m_x^2 - m_y^2} \log {m_y^2 \over m_x^2} + 2 \ .
\]

Likewise, we obtain the one-loop contributions from the bottom quark sector, $M^b$, as
%*********************  bottom and sbottom ******************************
\begin{eqnarray}
    M_{11}^b & = & {3 m_b^4 A_b^2 \Delta_{\tilde b}^2 \over 8 \pi^2  v^2 \cos^2 \beta}
{g(m_{{\tilde b}_1}^2, \ m_{{\tilde b}_2}^2) \over (m_{{\tilde b}_2}^2 - m_{{\tilde b}_1}^2)^2}
 + {3 m_b^4 A_b \Delta_{\tilde b} \over 4 \pi^2 v^2 \cos^2 \beta}
{\log (m_{{\tilde b}_2}^2 / m_{{\tilde b}_1}^2) \over (m_{{\tilde b}_2}^2 - m_{{\tilde b}_1}^2)} \cr
& &\mbox{} + {3 m_b^4 \over 8 \pi^2 v^2 \cos^2 \beta}
\log \left ( {m_{{\tilde b}_1}^2  m_{{\tilde b}_2}^2 \over m_b^4} \right ) \ , \cr
%%%%%%%%%
    M_{22}^b & = & {3 m_b^4 \lambda^2 s^2 \Delta_{\tilde b}^2 \over 8 \pi^2  v^2 \cos^2 \beta}
{g(m_{{\tilde b}_1}^2, \ m_{{\tilde b}_2}^2) \over (m_{{\tilde b}_2}^2 - m_{{\tilde b}_1}^2)^2}  \ , \cr
%%%%%%%%%
    M_{33}^b & = & {3 m_b^4 \lambda^2 \Delta_{\tilde b}^2 \over 8 \pi^2 \cot^2 \beta}
{g(m_{{\tilde b}_1}^2, \ m_{{\tilde b}_2}^2) \over (m_{{\tilde b}_2}^2 - m_{{\tilde b}_1}^2 )^2} \ , \cr
%%%%%%%%%
    M_{12}^b & = &\mbox{} - {3 m_b^4 \lambda A_b s \Delta_{\tilde b}^2
\over 8 \pi^2 v^2 \cos^2 \beta}
{g(m_{{\tilde b}_1}^2, \ m_{{\tilde b}_2}^2) \over (m_{{\tilde b}_2}^2 - m_{{\tilde b}_1}^2)^2}
- {3 m_b^4 \lambda s \Delta_{\tilde b} \over 8 \pi^2 v^2 \cos^2 \beta}
{\log (m_{{\tilde b}_2}^2 / m_{{\tilde b}_1}^2) \over (m_{{\tilde b}_2}^2 - m_{{\tilde b}_1}^2)} \cr
%%%%%%%%%
    M_{13}^b & = &\mbox{} - {3 m_b^4 \lambda A_b \Delta_{\tilde b}^2
\over 8 \pi^2 v \cos \beta \cot \beta}
{g(m_{{\tilde b}_1}^2, \ m_{{\tilde b}_2}^2) \over (m_{{\tilde b}_2}^2 - m_{{\tilde b}_1}^2)^2}
- {3 m_b^4 \lambda \sin \beta \Delta_{\tilde b} \over 8 \pi^2 v \cos^2 \beta}
{\log (m_{{\tilde b}_2}^2 / m_{{\tilde b}_1}^2) \over (m_{{\tilde b}_2}^2 - m_{{\tilde b}_1}^2) }  , \cr
%%%%%%%%%
    M_{23}^b & = & {3 m_b^4 \lambda^2 s \Delta_{\tilde b}^2 \over 8 \pi^2 v \cos \beta \cot \beta}
{g(m_{{\tilde b}_1}^2, \ m_{{\tilde b}_2}^2) \over (m_{{\tilde b}_2}^2 - m_{{\tilde b}_1}^2)^2 }
- {3 m_b^2 \lambda^2 s \tan \beta \over 8 \pi^2 v \cos \beta}
f(m_{{\tilde b}_1}^2, \ m_{{\tilde b}_2}^2) ,
\end{eqnarray}
where
\begin{eqnarray}
 \Delta_{\tilde b} & = & A_b - \lambda s \tan \beta  \  ,
\end{eqnarray}
and we obtain the one-loop contributions from the exotic quark sector, $M^k$, as
%*********************  exotic quark and squark ******************************
\begin{eqnarray}
    M_{11}^k & = & {3 m_k^4 \lambda^2 v^2 \sin^2 \beta \Delta_{\tilde k}^2 \over 8 \pi^2  s^2 }
{g(m_{{\tilde k}_1}^2, \ m_{{\tilde k}_2}^2) \over (m_{{\tilde k}_2}^2 - m_{{\tilde k}_1}^2)^2}  \ , \cr
%%%%%%%%%
    M_{22}^k & = & {3 m_k^4 \lambda^2 v^2 \cos^2 \beta \Delta_{\tilde k}^2 \over 8 \pi^2 s^2}
{g(m_{{\tilde k}_1}^2, \ m_{{\tilde k}_2}^2) \over (m_{{\tilde k}_2}^2 - m_{{\tilde k}_1}^2 )^2} \ , \cr
%%%%%%%%%
    M_{33}^k & = & {3 m_k^4 A_k^2 \Delta_{\tilde k}^2 \over 8 \pi^2 s^2}
{g(m_{{\tilde k}_1}^2, \ m_{{\tilde k}_2}^2) \over (m_{{\tilde k}_2}^2 - m_{{\tilde k}_1}^2)^2}
 + {3 m_k^4 A_k \Delta_{\tilde k} \over 4 \pi^2 s^2}
{\log (m_{{\tilde k}_2}^2 / m_{{\tilde k}_1}^2) \over (m_{{\tilde k}_2}^2 - m_{{\tilde k}_1}^2)} \cr
& &\mbox{} + {3 m_k^4 \over 8 \pi^2 s^2}
\log \left ( {m_{{\tilde k}_1}^2  m_{{\tilde k}_2}^2 \over m_k^4} \right ) \ , \cr
%%%%%%%%%
    M_{12}^k & = & {3 m_k^4 \lambda^2 v^2 \sin 2 \beta \Delta_{\tilde k}^2 \over 16 \pi^2 s^2}
{g(m_{{\tilde k}_1}^2, \ m_{{\tilde k}_2}^2) \over (m_{{\tilde k}_2}^2 - m_{{\tilde k}_1}^2)^2 }
- {3 m_k^2 \lambda^2 v^2 \sin 2 \beta \over 16 \pi^2 s^2}
f(m_{{\tilde k}_1}^2, \ m_{{\tilde k}_2}^2) , \cr
%%%%%%%%%
    M_{13}^k & = &\mbox{} - {3 m_k^4 \lambda A_k v \sin \beta \Delta_{\tilde k}^2
\over 8 \pi^2 s^2}
{g(m_{{\tilde k}_1}^2, \ m_{{\tilde k}_2}^2) \over (m_{{\tilde k}_2}^2 - m_{{\tilde k}_1}^2)^2} \cr
& &\mbox{}- {3 m_k^4 \lambda v \sin \beta \Delta_{\tilde k} \over 8 \pi^2 s^2}
{\log (m_{{\tilde k}_2}^2 / m_{{\tilde k}_1}^2) \over (m_{{\tilde k}_2}^2 - m_{{\tilde k}_1}^2)} \cr
%%%%%%%%%
    M_{23}^k & = &\mbox{} - {3 m_k^4 \lambda A_k v \cos \beta \Delta_{\tilde k}^2 \over 8 \pi^2 s^2}
{g(m_{{\tilde k}_1}^2, \ m_{{\tilde k}_2}^2) \over (m_{{\tilde k}_2}^2 - m_{{\tilde k}_1}^2)^2} \cr
& &\mbox{}- {3 m_k^4 \lambda v \cos \beta \Delta_{\tilde k} \over 8 \pi^2 s^2}
{\log (m_{{\tilde k}_2}^2 / m_{{\tilde k}_1}^2) \over (m_{{\tilde k}_2}^2 - m_{{\tilde k}_1}^2) }  ,
\end{eqnarray}
where
\begin{eqnarray}
 \Delta_{\tilde k} & = & A_k - \lambda v \tan \alpha \  .
\end{eqnarray}
Note that similar calculations have been performed for the case of explicit CP violation elsewhere
to obtain the mass matrix for the neutral Higgs bosons in the USSM, which depends on CP phases.
If the CP phases be zero, the above results are consistent with the previous article [32].

In order to obtain analytical expressions for the eigenvalues of $M$,
one has to solve a complicated cubic equation.
We sort the masses of the scalar Higgs bosons on increasing order
such that $m_{S_1} \le m_{S_2} \le m_{S_3}$, where $S_i$ ($i=1,2,3$) are
three neutral scalar Higgs bosons in the USSM, and $m_{S_i}$ ($i=1,2,3$) are their corresponding masses.
The orthogonal transformation matrix, $O$,
which is related to the eigenvalues of $M$ as ${\rm diag}(m_{S_1}^2, m_{S_2}^2, m_{S_3}^2) = O^T M O$,
with the orthogonal condition, $O^T O=1$.

The upper bound on the lightest scalar Higgs boson mass at the one-loop level may be
obtained analytically by observing that the smallest eigenvalue of a symmetric matrix is smaller
than the smaller eigenvalue of its upper-left $2 \times 2$ submatrix [8].
Thus, we have
\begin{eqnarray}
m_{S_1}^2 & \le & \lambda^2 v^2 \sin^2 2 \beta + m_Z^2 \cos^2 2 \beta
+ 2 {g'}^2 v^2 ({\tilde Q}_1 \cos^2 \beta + {\tilde Q}_2 \sin^2 \beta)^2 \cr
& &\mbox{} + {3 m_t^4 \over 8 \pi^2 v^2}
{\Delta_{\tilde t}^4
\over (m_{{\tilde t}_2}^2 - m_{{\tilde t}_1}^2)^2}
g(m_{{\tilde t}_1}^2, \ m_{{\tilde t}_2}^2) \cr
&  & \mbox{} + {3 m_t^4 \over 4 \pi^2 v^2}
{\Delta_{\tilde t}
\over (m_{{\tilde t}_2}^2 - m_{{\tilde t}_1}^2)}
\log \left({m_{{\tilde t}_2}^2 \over m_{{\tilde t}_1}^2}\right)
+ {3 m_t^4 \over 8 \pi^2 v^2}
\log({m_{{\tilde t}_1}^2 m_{{\tilde t}_2}^2 \over m_t^4}) \cr
& &\mbox{} + {3 m_b^4 \over 8 \pi^2 v^2}
{\Delta_{\tilde b}^4
\over (m_{{\tilde b}_2}^2 - m_{{\tilde b}_1}^2)^2}
g(m_{{\tilde b}_1}^2, \ m_{{\tilde b}_2}^2) \cr
&  & \mbox{} + {3 m_b^4 \over 4 \pi^2 v^2}
{\Delta_{\tilde b}
\over (m_{{\tilde b}_2}^2 - m_{{\tilde b}_1}^2)}
\log \left({m_{{\tilde b}_2}^2 \over m_{{\tilde b}_1}^2}\right)
+ {3 m_b^4 \over 8 \pi^2 v^2}
\log({m_{{\tilde b}_1}^2 m_{{\tilde b}_2}^2 \over m_b^4}) \cr
& &\mbox{} + {3 m_k^4 \lambda^2 v^2 \Delta_{\tilde k}^2 \over 8 \pi^2 s^2}
{g(m_{{\tilde k}_1}^2, \ m_{{\tilde k}_2}^2)
\over (m_{{\tilde k}_2}^2 - m_{{\tilde k}_1}^2)^2}
- {3 m_k^2 \lambda^2 v^2 \sin^2 2 \beta \over 16 \pi^2 s^2} f(m_{{\tilde k}_1}^2, \ m_{{\tilde k}_2}^2) \ ,
\end{eqnarray}
where the first three terms come from $V^0$ while the remaining terms come from $V^1$.
Note that there is no explicit appearance of the renormalization scale in the above formula for
the upper bound on the lightest scalar Higgs boson mass, but it may depend on the renormalization
scale implicitly through other relevant parameters.

%*****************************************************************
\subsection{HIGGS PRODUCTION}
%*****************************************************************

At the LHC, it is expected that signals for the SM Higgs boson would be produced copiously.
The dominant production mechanism for the SM Higgs boson is the gluon fusion process,
where a pair of gluons are fused into a triangular shaped fermion loop
that would eventually decay into the SM Higgs boson, $\phi$.
It may be described as
\[
pp \to gg \to \phi \ .
\]
The role of the fermion loop is to mediate the couplings between the gluon pair and the SM Higgs boson.
For the SM, the gluon fusion process is in effect activated by the top quark loops alone,
since the coupling between the SM Higgs boson and other fermions are negligible as compared
to the coupling between the SM Higgs boson and top quark.

In supersymmetric models, other fermions may take part in the gluon fusion process
to produce the scalar Higgs bosons.
For example, in the MSSM, the coupling between the lightest neutral scalar Higgs boson
and bottom quark can be strong for large $\tan \beta$, and thus the gluon fusion process
via bottom quark loop may be enhanced.

The USSM has additional fermions, that is, the exotic quarks.
In principle, the exotic quarks also may take part in the gluon fusion process.
In practice, it depends on their masses if the exotic quarks can play a dominant role or not.
For certain values for relevant parameters of the USSM,
the exotic quarks may have masses comparable to the electroweak scale,
and in this case the production of scalar Higgs bosons through the gluon fusion process
via the exotic quark loop may become important for Higgs search at the LHC.
Let us consider in more detail this possibility.

The gluon fusion process in the USSM at the LHC may also be described as
\[
pp \to gg \to S_j \ ,
\]
where $S_j$ ($j=1,2,3$) are the neutral scalar Higgs bosons in the USSM.
The relevant parts of interaction Lagrangian between the neutral scalar Higgs bosons and top,
bottom, and the exotic quark pairs are given as
\begin{eqnarray}
{\mathcal L}_b & = &\mbox{} - {g_2 m_b \over 2 m_W \cos \beta} O_{1j} S_j {\bar b} b \ , \cr
{\mathcal L}_t & = &\mbox{} - {g_2 m_t \over 2 m_W \sin \beta} O_{2j} S_j {\bar t} t \ , \cr
{\mathcal L}_k & = &\mbox{} - {m_k \over s} O_{3j} S_j {\bar k} k \ , \nonumber
\end{eqnarray}
where $O_{ij}$ are the elements of the orthogonal transformation matrix that diagonalizes $M$,
the mass matrix for the neutral scalar Higgs bosons.
Thus, the coupling coefficients of the neutral scalar Higgs bosons to relevant quark pairs are
easily obtained as
\begin{eqnarray}
G_{bbS_j} & = & {g_2 m_b O_{1j} \over 2 m_W \cos \beta} \ , \cr
G_{ttS_j} & = & {g_2 m_t O_{2j} \over 2 m_W \sin \beta} \ , \cr
G_{kkS_j} & = & {m_k \over s} O_{3j} \ . \nonumber
\end{eqnarray}
In terms of these coupling coefficients, the cross section for production
of the neutral scalar Higgs bosons at the LHC through the gluon fusion process
via relevant quark loops can be calculated.
In particular, we are interested in the exotic quarks.

Let us study the gluon fusion process where the exotic quark loops are involved.
The renormalization and factorization scales are taken to be the neutral scalar Higgs boson mass.
At parton level, two gluons are annihilated into a neutral scalar Higgs boson via the exotic quark loop.
The leading-order calculation for the gluon-gluon annihilation into $S_j$ ($j = 1,2,3$) yields
\begin{equation}
{\hat \sigma}_j({\hat s} ) = \sigma^k_j (gg \to S_j) = {\alpha_s^2(m_{S_j}) O_{3j}^2 \over 576 s^2 \pi}
    \delta \left (1-{m_{S_j}^2 \over {\hat s}^2 }\right )
    \left|{3 \over 2} \tau \left [1 + (1 - \tau) {\textsl f}(\tau) \right] \right|^2 \ ,
\end{equation}
where  ${\hat s}$ is the square of the c.m. energy of two gluons, $\alpha_s (m_{S_j})$ is
the strong coupling constant evaluated at $m_{S_j}$, $\tau = 4 m_k^2/m_{S_j}^2$ is the scaling variable,
and the function ${\textsl f}(\tau)$ is defined as
\begin{eqnarray}
{\textsl f}(\tau) = \left \{
\begin{array}{cl}
{\rm arcsin}^2(1/\sqrt{\tau}) \ ,    & \qquad \tau \geq 1 \ , \cr
-{1 \over 4} \left[ \log \left( {\displaystyle {1+\sqrt{1+ \tau} \over 1 - \sqrt{1- \tau}} } \right) -i \pi \right]^2 \ ,
& \qquad \tau < 1 \ .
\end{array}\right.
\end{eqnarray}

In order to calculate the production cross section of the neutral scalar Higgs bosons
in proton-proton collisions at the LHC, we need to fold ${\hat \sigma}_j ({\hat s})$
with the gluon distribution functions.
Thus, the desired cross section at hadronic level is obtained by integrating the cross section
at parton level over the gluon luminosity [44,45] as
\[
\sigma^k_j (m^2_{S_j}) = \sigma^k_j (pp \to S_j)
= \int_{m^2_{S_j} / E^2_{c.m.}}^1 d\tau \int_{\tau}^1 dx
{\hat \sigma}_j ({\hat s} = \tau E^2_{c.m.}) g (x, m_{S_j}^2) g (\tau/x,m_{S_j}^2)
\]
where $E_{c.m.} = 14$ TeV is the c.m. energy of the proton-proton system at the LHC,
$g(x,  m_{S_j}^2)$ is the gluon distribution function at the factorization scale $m_{S_j}^2$,
and $x$ is the momentum fraction.
Similarly, $g (\tau/x,m_{S_j}^2)$ is the gluon distribution function for the other gluon.
Note that $\sigma^k_j$ ($j = 1,2,3$) are explicit functions of $m_{S_j}^2$.
Also,  they depend directly upon $O_{3j}^2$.
Likewise, if we suitably replace exotic quarks by top quarks,
we may calculate $\sigma^t_j (m_{S_j}^2)$ ($j = 1,2,3$),
which are the cross sections of the $S_j$ productions in proton-proton collisions
through gluon fusion process via top quark loop.

%***********************************************************************
\section{NUMERICAL ANALYSIS}
%***********************************************************************

For numerical analysis, we need to set the values of relevant parameters in the USSM.
The masses of the third generation are set as $m_t$ = 175 GeV and $m_b$ = 4 GeV.
For the mass of the exotic quark with electric charges $\pm 1/3$,
the Run 1 data from Tevatron have set the lower bound as 190 GeV [46].
Recently, the search for long-lived charged massive particles at Tevatron Run I data
put a more stringent experimental lower bound of 180 GeV on the exotic quark mass
at the 95 \% confidence level [47].
We would like to set $m_k$ = 400 GeV for our analysis.

Since the effective $U(1)'$ charges ${\tilde Q}_i$ always go together
with the $U(1)'$ gauge coupling constant [33], let us define the modified $U(1)'$ charges
as $Q_i = g'_1 {\tilde Q}_i$.
They satisfy $\sum Q_i = 0$, which can easily be seen from the $U(1)'$ gauge invariance condition
$\sum {\tilde Q}_i = 0$.
The modified $U(1)'$ charges receive strong experimental constraints from the experimental bounds
on the extra gauge boson mass $m_{Z'}$ and on the mixing angle
between the two neutral gauge boson $|\alpha_{ZZ'}|$.
In this article, we set the values of the modified $U(1)'$ charges
as  $Q_1 = - 1$, $Q_2 = - 0.1$, and $Q_3 = 1.1$,
which are picked up from the experimentally allowed area in the ($Q_1, Q_2$)-plane
by using $m_{Z'} > 600$ GeV and $|\alpha_{ZZ'}| < 2 \times 10^{-3}$ [33].

The other parameters are set as $\tan\beta =10$, $\lambda = 0.5$, $s = m_Q = m_K = 500$ GeV,
and $A_t = 1000$ GeV.
We allow the pseudoscalar Higgs boson mass may vary within $200 \le m_A {\rm (GeV) } \le 1000$.
The numerical integrations are performed through the gauss integration function of the CERN program library.
For the gluon distribution functions, we use the PDF library of the CTEQ6M [40,41].
For the above set of parameter values, we have $m_{S_1} \approx 138$ GeV.
It is roughly stable against the variation in the pseudoscalar Higgs boson mass.
The masses of the other two neutral scalar Higgs bosons are estimated
to be $200 < m_{S_2} {\rm (GeV) } < 788$ and $793 < m_{S_3} {\rm (GeV) } < 1000$.
We note that the upper bound on $m_{S_1}$ is estimated to be about 156 GeV as
we explore reasonable ranges of the parameter space in the USSM with the SUSY breaking scale of 1 TeV.
This value of the upper bound on $m_{S_1}$ at the one-loop level may be
compared with that of the MSSM, where the upper bound on the mass of the lighter
one of two neutral scalar Higgs bosons is about 135 GeV.

Let us introduce for convenience $R^t_j$ and $R^k_j$ as
\[
R_j^t = \left ({O_{2j} \over \sin \beta} \right )^2 \ , \qquad R_j^k = O_{3j}^2 \ ,
\]
where $t$ and $k$ stand for top quark and the exotic quark, respectively.
These two coefficients are directly proportional to $\sigma^t_j (m_{S_j}^2)$
and $\sigma^k_j (m_{S_j}^2)$, respectively.
Also, they are proportional to the coupling coefficients of $S_j$
to top quarks and the exotic quarks.
Therefore, by analyzing the behavior of these coefficients,
we may study the possibility of the productions of the neutral scalar Higgs bosons
in the USSM at the LHC.

It is observed that $S_1$ couples strongly to a top quark pair if its mass is large.
Thus, for $m_{S_1}\approx 138$ in our case, we have $O_{21}^2 \approx 1$.
In an extreme case for example where $O_{21}^2 = 1$, one would have
\[
    O = \left ( \begin{array}{ccc} 0 & O_{12} & O_{13} \cr 1 & 0 &  0 \cr 0 & O_{32} & O_{33}
    \end{array} \right )
\]
where $O_{32}^2 + O_{33}^2 = 1$, due to the orthogonality of the transformation matrix $O$.
In this case, $R^t_1 = 1/\sin\beta$, and $R^t_2 = R^t_3 = 0$ as well as $R^k_1 = 0$,
whereas $R^k_2 + R^k_3 = 1$.
In other words, $\sigma^t_1$ would be large, and $\sigma^k_2 + \sigma^k_3$ would be comparably large.
However, $\sigma^t_2 \approx \sigma^t_3 \approx 0$ and $\sigma^k_1\approx 0$.

In Fig. 1, we plot $R^t_j$ and $R^k_j$ ($j = 1,2,3$) as functions of the scalar Higgs boson masses.
The figure is actually a composition of three separate pieces:
The left piece of the figure shows $R^t_1$ and $R^k_1$ as functions of $m_{S_1}$;
the middle one shows $R^t_2$ and $R^k_2$ as functions of $m_{S_2}$;
and the right one shows $R^t_3$ and $R^k_3$ as functions of $m_{S_3}$.
Since $m_{S_1} \approx 138$ GeV for the whole parameter ranges we consider,
$R^t_1$ and $R^k_1$ are shown as short vertical line segments.
We see that $R^t_1$ is effectively 1 whereas $R^k_1 \approx 10^{-3}$.

On the other hand, as $m_{S_2}$ varies from 200 GeV to 788 GeV,
the plots of $R^t_2$ and $R^k_2$ exhibit wide variations.
Notice in Fig. 1 that $R^k_2$ increases from about $10^{-3}$ to practically 1,
while $R^t_2$ decreases from about $10^{-2}$ to zero.
The cross over occurs at $m_{S_2} =560$ GeV.
That is, $R_2^k > R_2^t$ for $560 < m_{S_2} < 788$ GeV.
Similarly, as $m_{S_3}$ varies from 793 GeV to 1000 GeV, the plots of $R^t_3$
and $R^k_3$ also vary widely.
The size of $R^t_3$ stays below $10^{-2}$ while $R^k_3$ changes from 1 down
to about $10^{-2}$.
We see that $R^k_2 + R^k_3 \approx 1$, but one may be larger than the other,
depending on the values of parameters.
For the whole range of $m_{S_3}$, we have $R_3^k > R_3^t$.

In Fig. 2, we plot $\sigma^t_j$ and $\sigma^k_j$ ($j = 1,2,3$)
as functions of the scalar Higgs boson masses.
Fig. 1 and Fig. 2 look alike, since both $R$'s and $\sigma$'s depend essentially
on $O_{ij}$'s, but details are somewhat different.
Like Fig. 1, this figure is also a composition of three separate pieces.
The left piece of the figure shows $ \sigma^t_1$ and $\sigma^k_1$ as functions of $m_{S_1}$;
the middle one shows $\sigma^t_2$ and $\sigma^k_2$ as functions of $m_{S_2}$;
and the right one shows $\sigma^t_3$ and $\sigma^k_3$ as functions of $m_{S_3}$.

For $S_1$ productions at the LHC, the gluon fusion process via top quark loop is
definitely dominant over the process via the exotic quark loop,
for the whole parameter ranges we consider.
We see that $\sigma^t_1 \approx 10^4$ fb whereas $\sigma^k_1$ is
negligibly less than 2 fb for $m_{S_1} \approx 138$ GeV.
On the other hand, for $S_2$ and $S_3$ productions at the LHC, the gluon fusion process
via the exotic quark loop may be comparable to or more dominant over the process
via top quark loop, for some parameter ranges.
In Fig. 2, one can see that $\sigma^k_2 > \sigma^t_2$ for $710 < m_{S_2} < 788$ GeV.
The most interesting case is the $S_3$ productions.
For the whole parameter ranges we consider, i.e., for the whole range of $m_{S_3}$
from 793 GeV to 1000 GeV, we have $\sigma^k_3$ is larger than $\sigma^t_3$.
It is remarkable that both $\sigma^k_2$ and $\sigma^k_3$ may be as large as 0.05 pb,
depending on the parameter values.

The cross sections for $S_j$ productions in the USSM may be compared with the MSSM Higgs searches.
There are two neutral scalar Higgs bosons, denoted as $h$ and $H$, in the MSSM,
Some years ago, Spira and his colleagues have calculated the production cross sections of
these neutral scalar Higgs bosons through the gluon fusion process for $\tan \beta=1.5$
by including some QCD corrections to the production process and
by considering the two-loop corrections to the MSSM Higgs sector [40,41].
The result for the $h$ production cross section is 80 pb for $m_h \approx$ 80 GeV,
and the results for the $H$ production cross sections are 10 pb for $m_H \approx 200$ GeV,
0.02 pb for $m_H \approx 800$ GeV, and 0.05 pb for $m_H \approx 1000$ GeV.
Compared with these numbers, the results of our numerical analysis indicate that the LHC may
be possible to compare the predictions between the MSSM and the USSM,
and test experimentally the contributions of the exotic quarks
to the productions of the neutral scalar Higgs bosons in the USSM through the gluon fusion process.

One might argue that our numerical numerical analysis is based on a specific choice of parameters.
However, it may well be expected that the contribution of the exotic quarks to the production of one of the heavier
neutral scalar Higgs bosons may be equal to, or larger than, the contribution of the top quarks,
for a wide parameter space.
The stringent bounds on the $Z$-$Z'$ mixing require the parameter $s$ to be rather as large as 500 GeV.
Therefore, one of the neutral scalar Higgs bosons is predominantly SU(2)-singlet, and its coupling to top
quarks is suppressed.
Since, at the same time, we choose a large value for the Yukawa coupling of the Higgs singlet
to the exotic quarks as $h_k = 0.8$, it is not surprising at all that the exotic quarks give a sizeable
contribution to the production of the scalar Higgs boson that is predominantly singlet.

%*****************************************************************
\section{DISCUSSIONS}
%*****************************************************************

The main purpose of this article is to emphasize the exotic quark effects
on the scalar Higgs productions via the gluon fusion process.
For experimental examinations of these effects, the Higgs decay modes in the USSM need to be investigated in detail,
since the decay of the Higgs boson into a gluon pair is affected by the exotic quarks through their triangular loop.
Comprehensive investigations are under way.
Here, we would like to make a few preliminary remarks.
The Higgs decay into a fermion pair is usually more dominant if the fermion is heavier,
since the coupling of the Higgs bosons to a fermion pair is proportional to the fermion mass.
Thus, the Higgs decay into a pair of bottom quarks is most dominant
for a wide area of the relevant parameter space if the Higgs boson is
lighter than 130 GeV, and the Higgs decay into a pair of gauge bosons is
more dominant than the other decay modes if the Higgs boson is heavier than 130 GeV.

In the USSM, the coupling of the Higgs boson to a fermion pair also significantly depends on
$\tan \beta$ and $O_{ij}$, besides the
fermion mass.
In our numerical analysis, the Higgs decays into a fermion pair depend weakly on $\tan \beta$,
as we take an  intermediate value for it:$\tan \beta = 10$.
However, the dependence of the Higgs decays into a fermion pair on $O_{ij}$ still persists.
Since  $O_{21}^2 \approx 1$ and the coupling of the lightest scalar Higgs boson to a pair of bottom quarks is small,
$G_{bbS_1} \sim O_{11} \approx 0$, for the parameter region we consider,
the coupling of the lightest scalar Higgs boson to a pair of charm quarks would be stronger
than that to a pair of bottom quarks,
Therefore, in the parameter region we consider, where the mass of the lightest scalar Higgs boson is around 138 GeV,
the lightest scalar Higgs boson might dominantly decay into a pair of charm quarks or a pair of $W$ bosons.

The Higgs decays into a gluon pair is generally not negligible if the Higgs boson mass is around 138 GeV.
For the lightest scalar Higgs boson, the effect of the exotic quarks through the triangular loop is not significant at all,
since $R_1^k$ is $10^3$ times as small as $R_1^t$ as one can easily notice in Fig. 1.
Thus, the decay of the lightest scalar Higgs boson via the triangular loop of exotic quark into a gluon pair is negligible.
For the heavier scalar Higgs bosons, their decays into a gluon pair
via the triangular loop of the exotic quark is considerably
significant since $R_i^k$ ($i=2,3$) is comparable with $R_i^t$.

It is always important, as well as interesting, to distinguish the Higgs bosons
in the USSM from those in such models as the SM or the MSSM.
We expect that it is rather easy to distinguish the lightest scalar Higgs boson in the USSM from the SM Higgs boson,
in the parameter region we consider, since the decay patterns between them are very different.
The SM Higgs boson exhibits that it decays more dominantly into a pair of bottom quarks than into a pair of charm quarks.
On the other hand, the lightest scalar Higgs boson in the USSM shows an opposite decay pattern,
in the parameter region we consider.

However, distinguishing the lightest scalar Higgs boson in the USSM from the corrsponding one
in the MSSM might be very difficult,
if their decay patterns are similar.
In this case, the search for the heavier Higgs bosons would be useful in order to distinguish the USSM from the MSSM.
In particular, as Fig. 1 shows, the decay modes of the heaviest scalar Higgs boson
via the triangular loop of exotic quarks into a gluon pair in the USSM might be strong
because $R_2^k$ is comparable with $R_2^t$.

The qualitative results of our analysis remains the same if we choose other parameter sets.
To be specific,  let us take $Q_1 = 0.9$ and $Q_2 = 0.025$ for the modified $U(1)'$ charges
of the Higgs doublets and singlet,
instead of $Q_1 = - 1$ and $Q_2 = - 0.1$ ($Q_3 = -Q_1 -Q_2$) which lead to Fig. 1 and Fig. 2,
These values for $Q_1$ and $Q_2$ are consistent with the experimental constraints
on both $\alpha_{ZZ'}$ and $m_{Z'}$,
for $s=500$ GeV and $\tan \beta = 10$.

For these values of $Q_1 = 0.9$ and $Q_2 = 0.025$, where the values of other parameters are
the same as in Figs. 1 and 2, we obtain
$R_i^q$ and $\sigma_i^q$ ($i=1,2,3; q=t,k$), which are plotted respectively in Figs. 3 and 4
as functions of the scalar Higgs boson masses.
One may notice that the exotic quark as light as 400 GeV might be able to play an important role
in the productions of the heavier scalar Higgs bosons via the gluon fusion process.

%*****************************************************************
\section{CONCLUSIONS}
%*****************************************************************

The Higgs sector of the USSM is studied at the one-loop level
by taking into account radiative corrections due to top, bottom, and the exotic quarks,
and their superpartners, in order to examine the possibility
of discovering one of the neutral scalar Higgs bosons in the USSM at the LHC.
A characteristic difference of the USSM from the MSSM is that there are exotic quarks in the USSM.
If the masses of these exotic quarks are a few hundred GeV, it would be difficult
to see direct signals of them at the LHC.
However, they may manifest themselves as intermediate states
in the production processes of other particles.
A clear candidate may be the gluon fusion process for the production of the neutral scalar Higgs bosons,
where fermion loops are engaged as intermediate states.
In the SM, as well as in the MSSM, the gluon fusion process is predominantly activated by a top quark loop.
In the USSM, it is possible that the exotic quark loop may also be involved,
and the contributions of the exotic quark loop to the gluon fusion process may
be significant as compared to the contributions due to top quark loop.

Indeed, the possibility depends on the values of the relevant parameters of the USSM.
The parameter region we explore are $\tan\beta =10$, $\lambda = 0.5$,
$s = m_Q = m_K = 500$ GeV, and $A_t = 1000$ GeV.
The mass of the pseudoscalar Higgs boson of the USSM, $m_A$,
is allowed to vary from 200 GeV to 1000 GeV, and the mass of the exotic quark $m_k$ is set as 400 GeV.
All parameters and VEVs are assumed to be real.
In this parameter region, the masses of the three neutral scalar Higgs bosons of the USSM
at the one-loop level are estimated to be $m_{S_1} \approx 138$ GeV,
$200 < m_{S_2} {\rm (GeV) } < 788$ and $793 < m_{S_3} {\rm (GeV) } < 1000$.
These masses lie within the experimentally allowed bound set by the LEP II data as well as the Tevatron data [48-50].

The result of our analysis on the productions of these neutral scalar Higgs bosons shows
that the exotic quarks in the USSM are competitively important for the gluon fusion process.
For $S_1$ productions, the gluon fusion process via top quark loop yields
exceedingly larger cross sections than the one via the exotic quark loop,
for the parameter region we consider.
However, we find that, for $S_2$ productions, the gluon fusion process
via the exotic quark loop may yield larger cross sections than the one via top quark loop,
when $S_2$ becomes more massive than 710 GeV.
For $S_3$ productions, exotic quarks are more important than top quark for the gluon fusion process,
for the entire parameter region we explore.

In conclusion, the exotic quarks in the USSM may help the neutral scalar Higgs bosons
in the same model to be discovered through gluon fusion process at the LHC.
To discover a neutral scalar Higgs boson at the LHC and identify
it as the USSM one would be very supporting evidences for supersymmetry and $E_6$ gauge group,
as well as a strong constraint upon the parameter space of the USSM.

%*****************************************************************
\vskip 0.3 in
\noindent
{\large {\bf ACKNOWLEDGMENTS}}
\vskip 0.2 in
This work is partly supported by KOSEF through a grant provided by the MOST in 2007
(project No. K2071200000107 A020000110) and by KOSEF through CHEP, Kyungpook National University.
One of the authors (PK) is grateful to KEK Theory Group where a part of this work has been performed.
The authors would like to acknowledge the support from KISTI (Korea Institute of Science
and Technology Information) under "The Strategic Supercomputing Support Program"
with Dr. Kihyeon Cho as the technical supporter.
The use of the computing system of the Supercomputing Center is also greatly appreciated.

%\vskip 0.3 in
%\vfil\eject

%******************************************************************

\vfil\eject
%***********************************************************************
{\large {\bf FIGURE CAPTION}}
\vskip 0.3 in
\noindent
FIG. 1. : The plot of $R_i^q$ ($i=1,2,3; q=t,k$) as functions of neutral scalar Higgs boson masses in the USSM.
The figure is a composition of three separate pieces:
The left piece of the figure shows $R^t_1$ and $R^k_1$ as functions of $m_{S_1}$;
the middle one shows $R^t_2$ and $R^k_2$ as functions of $m_{S_2}$;
and the right one shows $R^t_3$ and $R^k_3$ as functions of $m_{S_3}$.
The relevant parameters are set as $\tan\beta =10$, $\lambda = 0.5$, $s = m_Q = m_K = 500$ GeV, and $A_t = 1000$ GeV, $m_k = 400$ GeV, and $200 \le m_A {\rm { (GeV) }} \le 1000$.
The modified $U(1)'$ charges of the Higgs doublets and singlet are taken as $Q_1 = - 1$, $Q_2 = - 0.1$, and $Q_3 = 1.1$.

\vskip 0.3 in
\noindent
FIG. 2. : The plot of $\sigma_i^q$ ($i=1,2,3; q=t,k$) as functions of neutral scalar Higgs boson masses.
The figure is a composition of three separate pieces:
The left piece of the figure shows $ \sigma^t_1$ and $\sigma^k_1$ as functions of $m_{S_1}$;
the middle one shows $\sigma^t_2$ and $\sigma^k_2$ as functions of $m_{S_2}$;
and the right one shows $\sigma^t_3$ and $\sigma^k_3$ as functions of $m_{S_3}$.
The values for the relevant parameters are the same as Fig. 1.

\vskip 0.3 in
\noindent
FIG. 3. : The plot of $R_i^q$ ($i=1,2,3; q=t,k$) as functions of neutral scalar Higgs boson masses in the USSM.
The figure is a composition of three separate pieces:
The left piece of the figure shows $R^t_1$ and $R^k_1$ as functions of $m_{S_1}$;
the middle one shows $R^t_2$ and $R^k_2$ as functions of $m_{S_2}$;
and the right one shows $R^t_3$ and $R^k_3$ as functions of $m_{S_3}$.
The relevant parameters are the same as Fig. 1 except for  $Q_1 = 0.9$ and $Q_2 = 0.025$.
\vskip 0.3 in
\noindent
FIG. 4. : The plot of $\sigma_i^q$ ($i=1,2,3; q=t,k$) as functions of neutral scalar Higgs boson masses.
The figure is a composition of three separate pieces:
The left piece of the figure shows $ \sigma^t_1$ and $\sigma^k_1$ as functions of $m_{S_1}$;
the middle one shows $\sigma^t_2$ and $\sigma^k_2$ as functions of $m_{S_2}$;
and the right one shows $\sigma^t_3$ and $\sigma^k_3$ as functions of $m_{S_3}$.
The relevant parameters are the same as Fig. 1 except for $Q_1 = 0.9$ and $Q_2 = 0.025$.

\vfil\eject
%***********************************************************************
%Figure
\setcounter{figure}{0}
\def\figurename{}{}%
% (FIG 1)
\renewcommand\thefigure{FIG. 1}
\begin{figure}[t]
\begin{center}
\includegraphics[scale=0.6]{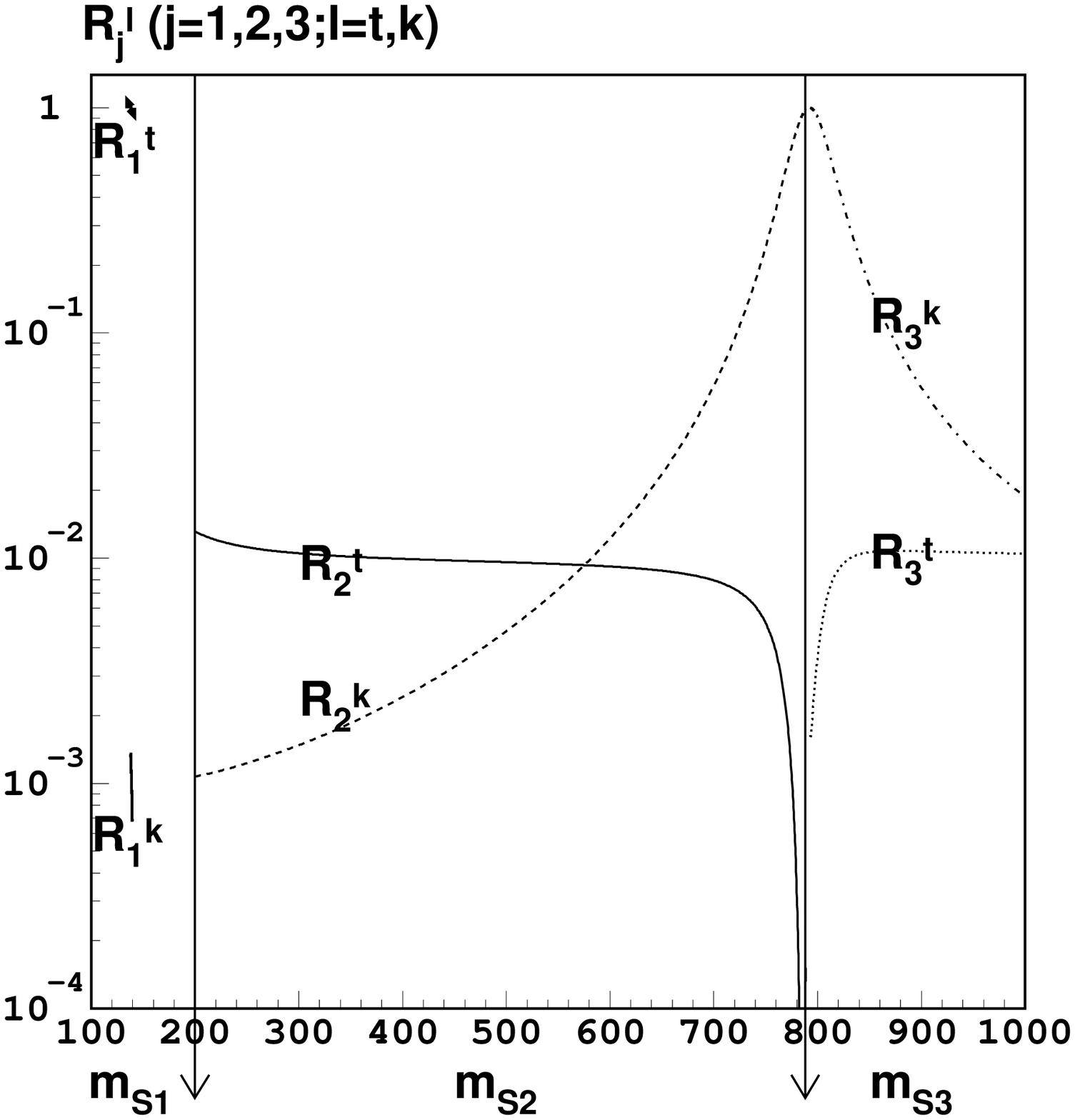}
\caption[plot]{The plot of $R_i^q$ ($i=1,2,3; q=t,k$) as functions of neutral scalar Higgs boson masses in the USSM.
The figure is a composition of three separate pieces:
The left piece of the figure shows $R^t_1$ and $R^k_1$ as functions of $m_{S_1}$;
the middle one shows $R^t_2$ and $R^k_2$ as functions of $m_{S_2}$;
and the right one shows $R^t_3$ and $R^k_3$ as functions of $m_{S_3}$.
The relevant parameters are set as $\tan\beta =10$, $\lambda = 0.5$, $s = m_Q = m_K = 500$ GeV, and $A_t = 1000$ GeV, $m_k = 400$ GeV, and $200 \le m_A {\rm { (GeV) }} \le 1000$.
The modified $U(1)'$ charges of the Higgs doublets and singlet are taken as $Q_1 = - 1$, $Q_2 = - 0.1$, and $Q_3 = 1.1$.}
\end{center}
\end{figure}

% (FIG 2)
\renewcommand\thefigure{FIG. 2}
\begin{figure}[t]
\begin{center}
\includegraphics[scale=0.6]{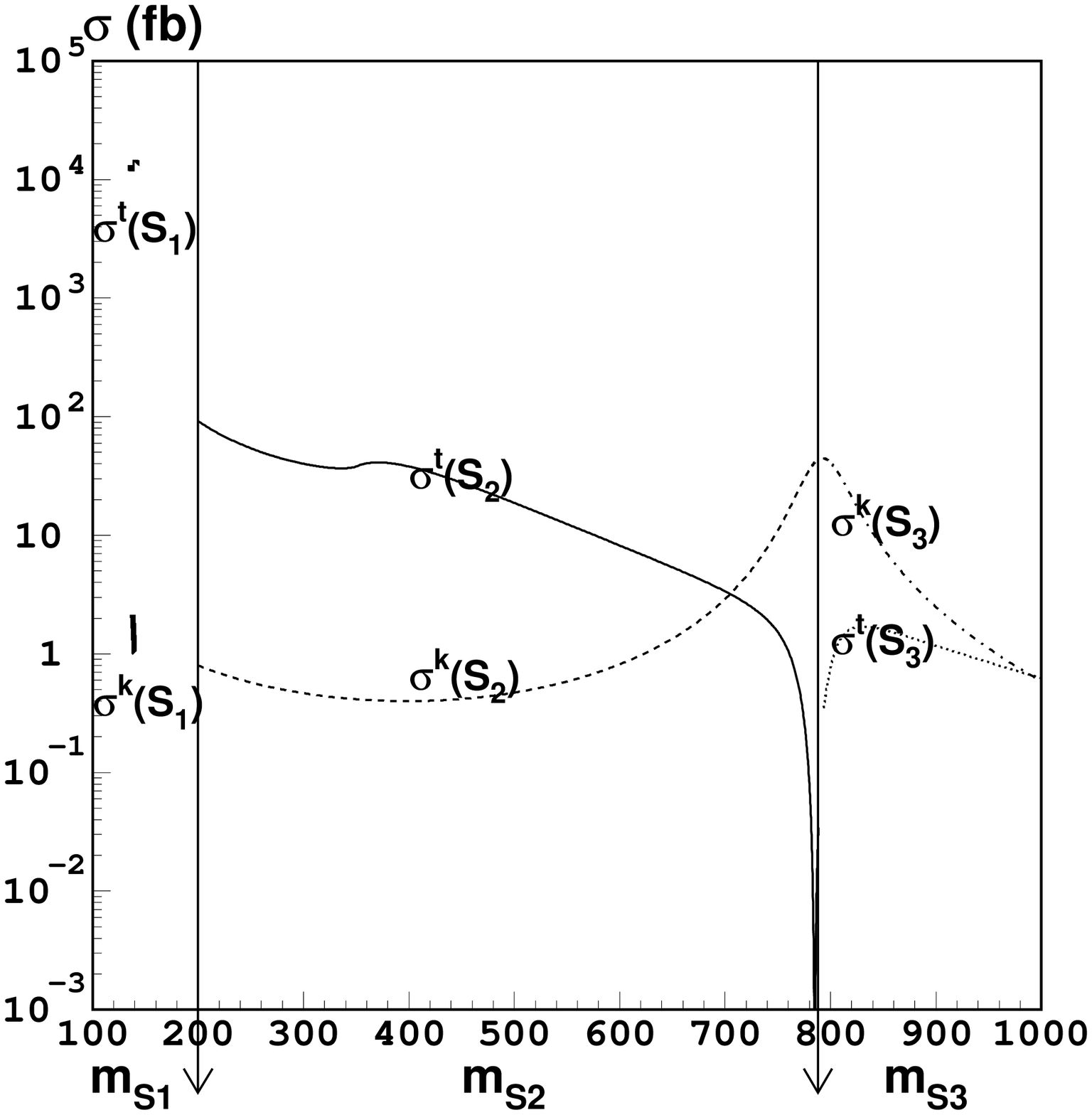}
\caption[plot]{ The plot of $\sigma_i^q$ ($i=1,2,3; q=t,k$) as functions of neutral scalar Higgs boson masses.
The figure is a composition of three separate pieces:
The left piece of the figure shows $ \sigma^t_1$ and $\sigma^k_1$ as functions of $m_{S_1}$;
the middle one shows $\sigma^t_2$ and $\sigma^k_2$ as functions of $m_{S_2}$;
and the right one shows $\sigma^t_3$ and $\sigma^k_3$ as functions of $m_{S_3}$.
The values for the relevant parameters are the same as Fig. 1.}
\end{center}
\end{figure}

% (FIG 3)
\renewcommand\thefigure{FIG. 3}
\begin{figure}[t]
\begin{center}
\includegraphics[scale=0.6]{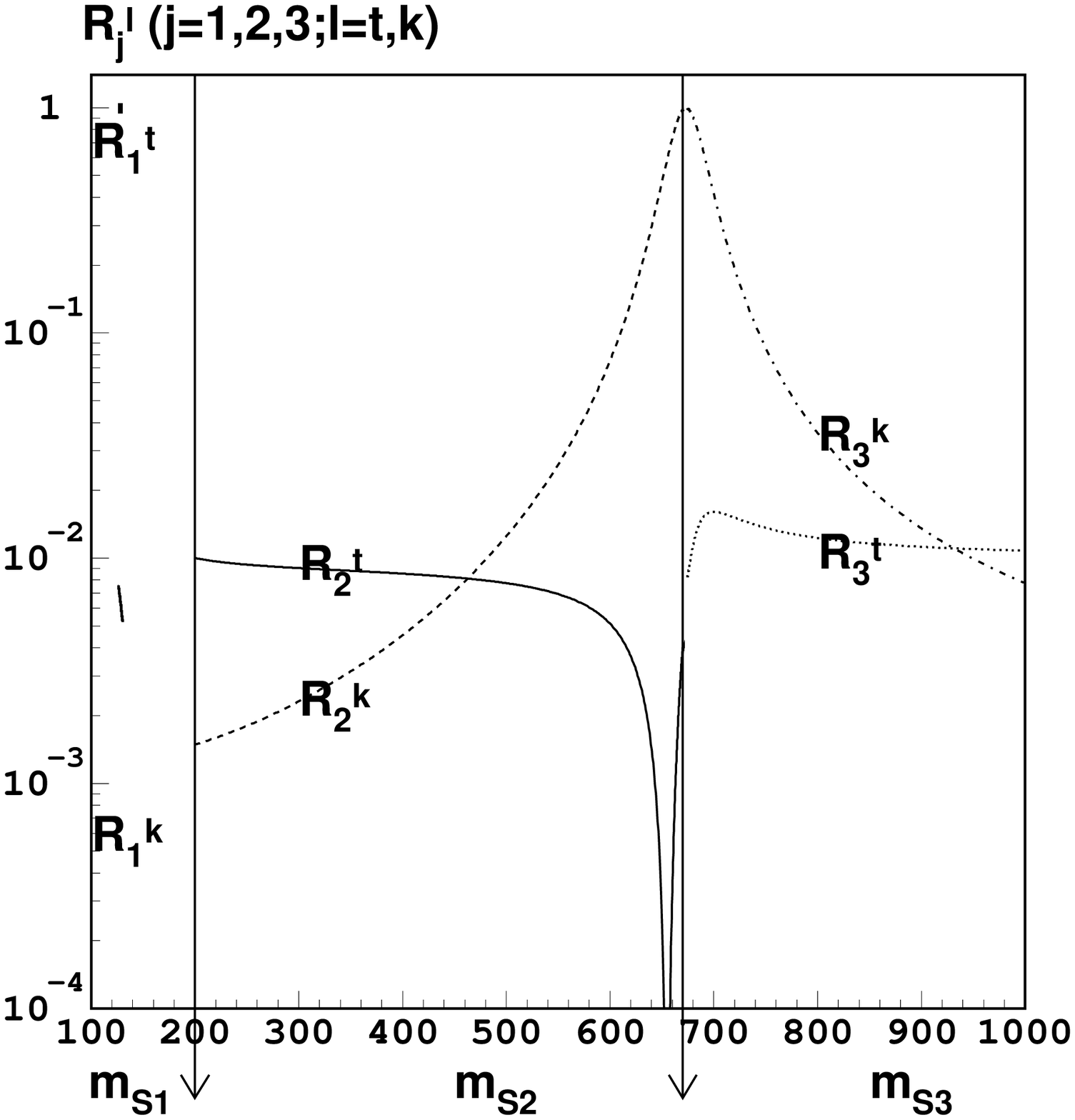}
\caption[plot]{The plot of $R_i^q$ ($i=1,2,3; q=t,k$) as functions of neutral scalar Higgs boson masses in the USSM.
The figure is a composition of three separate pieces:
The left piece of the figure shows $R^t_1$ and $R^k_1$ as functions of $m_{S_1}$;
the middle one shows $R^t_2$ and $R^k_2$ as functions of $m_{S_2}$;
and the right one shows $R^t_3$ and $R^k_3$ as functions of $m_{S_3}$.
The relevant parameters are the same as Fig. 1 except for  $Q_1 = 0.9$ and $Q_2 = 0.025$.}
\end{center}
\end{figure}

% (FIG 4)
\renewcommand\thefigure{FIG. 4}
\begin{figure}[t]
\begin{center}
\includegraphics[scale=0.6]{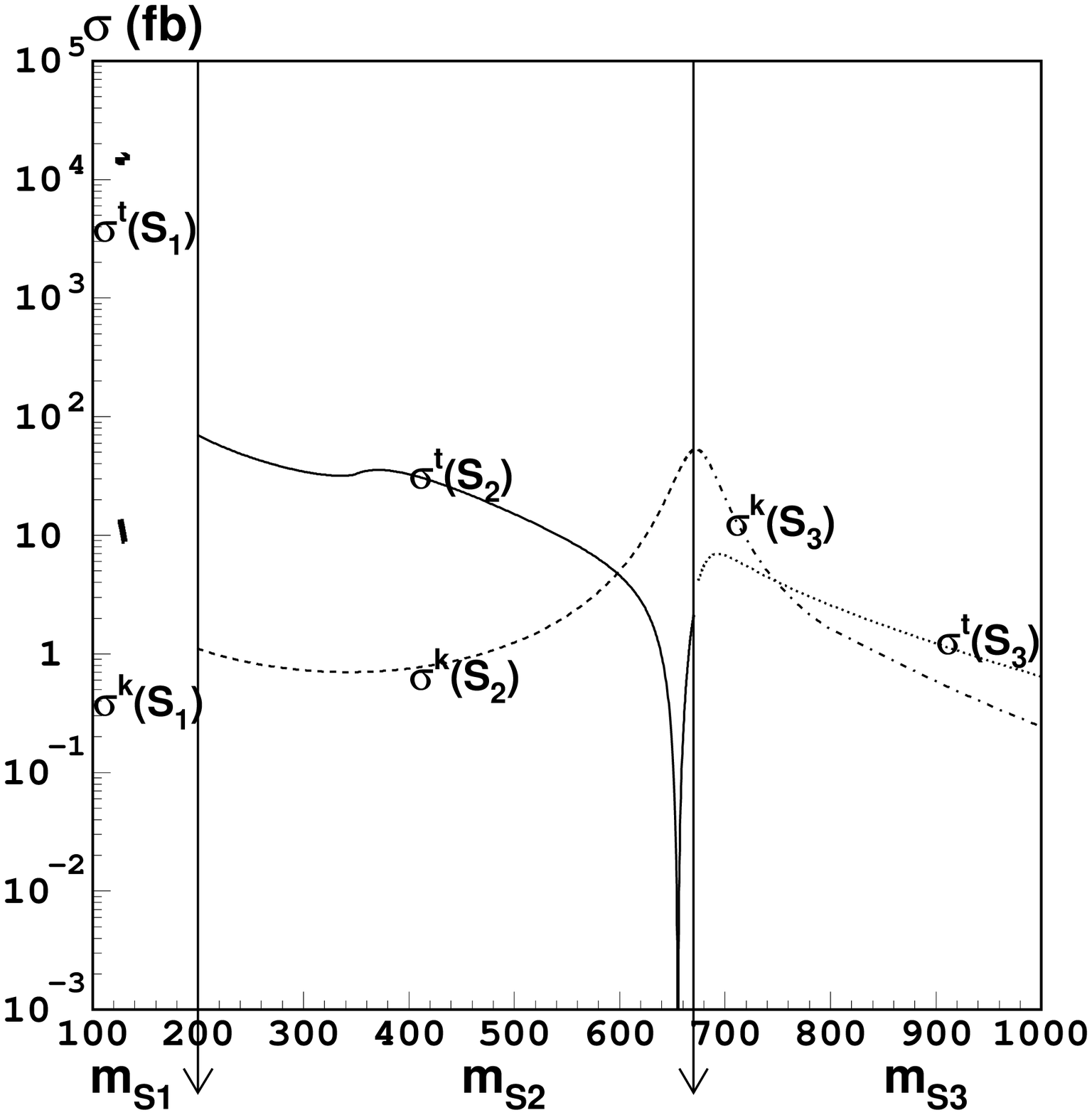}
\caption[plot]{ The plot of $\sigma_i^q$ ($i=1,2,3; q=t,k$) as functions of neutral scalar Higgs boson masses.
The figure is a composition of three separate pieces:
The left piece of the figure shows $ \sigma^t_1$ and $\sigma^k_1$ as functions of $m_{S_1}$;
the middle one shows $\sigma^t_2$ and $\sigma^k_2$ as functions of $m_{S_2}$;
and the right one shows $\sigma^t_3$ and $\sigma^k_3$ as functions of $m_{S_3}$.
The relevant parameters are the same as Fig. 1 except for  $Q_1 = 0.9$ and $Q_2 = 0.025$.}
\end{center}
\end{figure}

%***********************************************************************

\begin{thebibliography}{99}
%******************************************************************
\bibitem{1} P. Fayet and S. Ferrara, Phys. Rep. {\bf 32}, 249 (1977);
    P. Fayet, Phys. Rep. {\bf 105}, 21 (1984).
\bibitem{2} H. P. Nilles, Phys. Rep. {\bf 110}, 1 (1984).
\bibitem{3} J. F. Gunion, H.E. Haber, G.L. Kane, and S. Dawson,
    {\it The Higgs Hunters' Guide} (Addison-Wesley Redwood City, CA, 1990).
\bibitem{4} J. E. Kim and H. P. Nilles, Phys. Lett. B {\bf 138}, 150 (1984).
\bibitem{5} P. Fayet, Nucl. Phys. {\bf 90}, 104 (1975); Phys. Lett. B {\bf 69}, 489 (1977).
\bibitem{6} E. Cremmer, P. Fayet, and L. Girardello, Phys. Lett. B {\bf 122}, 41 (1983);
P. Fayet, Phys. Lett. B {\bf 125}, 178 (1983).
\bibitem{7} J. Ellis, J. F. Gunion, H. E. Haber, L. Roszkowski, and F. Zwirner, Phys. Rev. D {\bf 39}, 844, (1989).
\bibitem{8} M. Drees, Int. J. Mod. Phys. A {\bf 4}, 3635 (1989).
\bibitem{9} J. R. Espinosa and M. Quiros, Phys. Lett. B {\bf 279}, 92 (1992); Phys. Lett. B {\bf 302}, 51 (1993).
\bibitem{10} U. Ellwanger, Phys. Lett. B {\bf 303} (1993) 271.
\bibitem{11} P. N. Pandita, Z. Phys. C {\bf 59}, 575 (1993).
\bibitem{12} T. Elliott, S. F. King, and P. L. White, Phys. Rev. D {\bf 49} (1994) 2435;
S. F. King and P. L. White, Phys. Rev. D {\bf 52} (1995) 4183.
\bibitem{13} B. Ananthanarayan and P. N. Pandita, Phys. Lett. B {\bf 353}, 70 (1995);
Phys. Lett. B {\bf 371}, 245 (1996); Int. J. Mod. Phys. A {\bf 12}, 2321 (1997).
\bibitem{14} S. W. Ham, S. K. Oh, and B. R. Kim, J. Phys. G {\bf 22}, 1575 (1996);
Phys. Lett. B {\bf 383}, 179 (1996).
\bibitem{15} U. Ellwanger and C. Hugonie, Comput. Phys. Commun. {\bf 175}, 290 (2006);
Mod. Phys. Lett. A {\bf 22}, 1581 (2007).
\bibitem{16} C. Panagiotakopoulos and K. Tamvakis, Phys. Lett. B {\bf 446}, 224 (1999);
Phys. Lett. B {\bf 469}, 145 (1999).
\bibitem{17} C. Panagiotakopoulos and A. Pilaftsis, Phys. Rev. D {\bf 63}, 055003 (2001).
\bibitem{18} A. Dedes, C. Hugonie, S. Moretti, and K. Tamvakis, Phys. Rev. D {\bf 63}, 055009 (2001).
\bibitem{19} C. Hugonie, J. C. Romao, and A. M. Teixeira, JHEP {\bf 06}, 020 (2003).
\bibitem{20} S. W. Ham, S. K. Oh, C. M. Kim, E. J. Yoo, and D. Son, Phys. Rev. D {\bf 70}, 075001 (2004);
S. W. Ham, J. O. Im, and S. K. Oh, arXiv:hep-ph/0805.1115.
\bibitem{21} J. L. Hewett and T. G. Rizzo, Phys. Rep. {\bf 183}, 193 (1989).
\bibitem{22} A. Leike, Phys. Rep. {\bf 317}, 143 (1999).
\bibitem{23} M. Cvetic and P. Langacker, Phys. Rev. D {\bf 54}, 3570 (1996).
\bibitem{24} M. Cvetic, D. A. Demir, J. R. Espinosa, L. Everett, and P. Langacker,
             Phys. Rev. D {\bf 56}, 2861 (1997); Erratum-ibid. D {\bf 58}, 119905 (1998).
\bibitem{25} P. Langacker and J. Wang, Phys. Rev. D {\bf 58}, 115010 (1998)
\bibitem{26} D. A. Demir and N. K. Pak, Phys. Rev. D {\bf 57}, 6609 (1998);
\bibitem{27} Y. Daikoku and D. Suematsu, Phys. Rev. D {\bf 62}, 095006 (2000);
\bibitem{28} H. Amini, New J. Phys. {\bf 5}, 49 (2003).
\bibitem{29} S. F. King, S. Moretti, and R. Nevzorov, Phys. Rev. D {\bf 73}, 035009 (2006); Phys. Lett. B {\bf 634}, 278 (2006).
\bibitem{30} D. A. Demir and L. L. Everett, Phys. Rev. D {\bf 69}, 015008 (2004).
\bibitem{31} J. Erler, Nucl. Phys. {\bf B586}, 73 (2000).
\bibitem{32} S. W. Ham, E. J. Yoo, and S. K. Oh, Phys. Rev. D {\bf 76}, 015004 (2007).
\bibitem{33} S. W. Ham, E. J. Yoo, and S. K. Oh, Phys. Rev. D {\bf 76}, 075011 (2007);
S. W. Ham and S. K. Oh, Phys. Rev. D {\bf 76}, 095018 (2007).
\bibitem{34} V. Barger , P. Langacker, H. S. Lee, G. Shaughnessy, Phys. Rev. D {\bf 73}, 115010 (2006).
\bibitem{35} V. Barger, P. Langacker, H. S. Lee and G. Shaughnessy, Phys. Rev. D {\bf 73}, 115010 (2006).
\bibitem{36} J. Kang, P. Langacker, and B.D. Nelson, Phys. Rev. D {\bf 77}, 035003 (2008).
\bibitem{37} Taeil Hur, H. S. Lee, and S. Nasri, Phys. Rev. D {\bf 77}, 015008 (2008).
\bibitem{38} P. Langacker, arXiv:hep-ph/0801.1345.
\bibitem{39} H. M. Georgi, S. L. Glashow, M. E. Machacek, D. V. Nanopoulos, Phys. Rev. Lett. {\bf 40}, 692 (1978).
\bibitem{40} M. Spira, A. Djouadi, D. Graudenz, P. M. Zerwas, Nucl. Phys. B {\bf 453}, 17 (1995).
\bibitem{41} M. Spira, Fortsch. Phys. {\bf 46}, 203 (1998).
\bibitem{42} R. V. Harlander, Phys. Lett. B {\bf 492}, 74 (2000).
\bibitem{43} S. Coleman and E. Weinberg, Phys. Rev. D {\bf 7}, 1888 (1973).
\bibitem{44} H. L. Lai, J. Botts, J. Huston, J. G. Morfin, J. F. Owens, J. W. Qiu, W. K. Tung, H. Weerts,
Phys. Rev. D {\bf 51}, 4763 (1995).
\bibitem{45} J. Pumplin, D. R. Stump, J. Huston, H. L. Lai, P. Nadolsky, and W. K. Tung, JHEP 0207, 012 (2002);
D. Stump, J. Huston, J. Pumplin, W. K. Tung, H. L. Lai, S. Kuhlmann, and J. F. Owens,
JHEP 0310, 046 (2003).
\bibitem{46} CDF Collaboration, T. Affolder, {\it et al}., Phys. Rev. Lett. {\bf 84}, 835 (2000).
\bibitem{47} CDF Collaboration, D. Acosta, {\it et al}., Phys. Rev. Lett. {\bf 90}, 131801 (2003).
\bibitem{48} M. Tytgat, Proceedings of Moriond 2008 on Electroweak interactions and Unified Theories.
\bibitem{49} L. Zivkovic, Proceedings of Moriond 2008 on Electroweak interactions and Unified Theories.
\bibitem{50} K. Yorita, Proceedings of Moriond 2008 on Electroweak interactions and Unified Theories.

\end{thebibliography}
\end{document}